\documentclass[preprint,12pt]{elsarticle}
\usepackage{amssymb}
\usepackage{amsthm}
\usepackage[svgnames]{xcolor}
\usepackage{amsmath}

\theoremstyle{definition}
\newtheorem{definition}{Definition} 
\newtheorem{example}{Example}

\theoremstyle{plain}
\newtheorem{theorem}{Theorem} 
\theoremstyle{remark}

\newtheorem{claim*}{Claim}[theorem]

\usepackage[euler]{textgreek}
\newcommand{\str}[1]{\texttt{#1}}
\newcommand{\vvv}{\;\vert\;}
\newcommand{\eee}{\texttt{\textepsilon}}
\usepackage{pbox}

\journal{Information Processing Letters}

\begin{document}

\begin{frontmatter}

\begingroup
\let\newpage\relax
\title{A note on the class of languages generated by F-systems over regular languages\tnoteref{t1}}
\tnotetext[t1]{This is the accepted version of a manuscript published in {\it Information Processing Letters}, https://doi.org/10.1016/j.ipl.2022.106283, and it is made available under a CC-BY-NC-ND 4.0 license https://creativecommons.org/licenses/by-nc-nd/4.0/.
}

\author{Jorge C. Lucero}

\address{Dept.\ Computer Science, University of Bras\'{i}lia, Brasília DF 70910-900, Brazil.}

\ead{lucero@unb.br}

\author{S\l{}awek Staworko}

\address{Univ.\ Lille, CNRS, Inria, Centrale Lille, UMR 9189 - CRIStAL, F-59000 Lille,
France}

\ead{slawomir.staworko@univ-lille.fr}

\begin{abstract}
An F-system is a computational model that performs a folding operation on words of a given language, following directions coded on words of another given language. This paper considers the case in which both given languages are regular, and it shows that the class of languages generated by such F-systems is a proper subset of the class of linear context-free languages.
\end{abstract}

\endgroup
\end{frontmatter}

\section{Introduction}

\label{intro}

Geometric folding processes are ubiquitous in nature and technology, from the shaping of protein molecules \cite{Dobson2003} and the folding of leaves and insect wings \cite{Mahadevan2005}, to self-assembling robots \cite{Felton2014} and foldable space telescopes \cite{Lang2007}. In current days, it is usual to designate such processes under the general term of  ``origami'', in reference to the Japanese traditional art of creating figures by folding a sheet of paper \cite{Demaine2007}. 

From the perspective of the theory of formal languages, origami has been modeled by a word folding operation, which reorders symbols of a given word according to directions coded in another one \cite{Sburlan2011}. Using the folding operation, a folding system (F-system) of the form $\Phi=(L_1, L_2)$ may be defined, where $L_1$ (the core language) is the language that contains the words to be folded, and $L_2$ (the folding procedure language) is the language that contains words with the folding directions. Although this model is restricted to one dimensional folding and does not capture actual origami (i.e., on a bidimensional sheet), it may still be applied to characterize folding processes in molecular or DNA computing and related areas \cite{Kari2008, Rothemund2006, Rozenberg2001}.

The computing power of F-systems has been investigated by comparison with standard language classes
from the Chomsky hierarchy (i.e., regular, context-free, context-sensitive, recursive
and recursively enumerable languages). More recently  \cite{Lucero2019}, necessary conditions for a language to belong to classes generated when the core and the folding procedure languages are regular or context-free have been proposed in the form of pumping lemmas, similar to the well known pumping lemmas
for regular and context-free languages.

The present paper considers the case in which both the core and the folding procedure languages are regular. It has been demonstrated that the class of languages generated by such F-systems surpasses and strictly contains the regular languages \cite{Sburlan2011}. Here, it will be shown that the F-system class is a proper subset of the class of the linear context-free languages.

\section{Definitions}
\label{fsystems}

Let us first review the definitions of folding operations and systems \cite{Lucero2019}.

\begin{definition}
\label{def1}
Let $\Sigma$ be an alphabet, $\Gamma=\{\str{u}, \str{d}\}$, and $f:\Sigma^*\times\Sigma\times\Gamma\rightarrow\Sigma^*$ a function such that
\begin{equation*}
f(x, a, b)= \begin{cases}
ax & \text{if $b=\str{u}$},\\
xa & \text{if $b=\str{d}$}.
\end{cases}
\end{equation*}
Then, the folding function $h:\Sigma^*\times\Gamma^*\rightarrow\Sigma^*$ is a partial function defined by
\begin{equation*}
h(w, v)=
\begin{cases}
\eee & \text{if $|w|=|v|=0$},\\
f(h(w', v'), a, b) & \text{if $|w|=|v|>0$, with $w=w'a$, $v=v'b$},\\
\text{undefined} & \text{if $|w|\neq|v|$}.\rlap{\hspace*{140pt}\qed}
\end{cases}
\end{equation*}
\end{definition}

The computation of $h(w, v)$ may be regarded as a folding operation that rearranges the symbols of $w$. Words over $\Gamma$ describe how each folding must be performed, where symbol \str{u} represents a ``folding up'' action and symbol \str{d} represents a ``folding down'' action (see \cite{Lucero2019} for an illustration of the folding mechanism).

\begin{definition}
\label{fsystem}
A folding system (F-system) is a pair $\Phi=(L_1, L_2)$, where $L_1\subseteq \Sigma^*$ is the core language, and $L_2\subseteq \Gamma^*$ is the folding procedure language. The language of $\Phi$ is
\begin{equation*}
  L(\Phi)=\{h(w, v)|\,w\in L_1, v\in L_2, |w|=|v|\}.
  \tag*{\qed}
\end{equation*}
\end{definition}

\begin{definition}
The class of all languages generated by F-systems with core languages of a class $\mathcal{C}$ and folding procedure languages of a class $\mathcal{H}$ is
\begin{equation*}
  \mathcal{F}(\mathcal{C}, \mathcal{H})=\{L(\Phi)|\,\Phi=(L_1, L_2), L_1\in\mathcal{C}, L_2\in\mathcal{H}\}.
  \tag*{\qed}
\end{equation*}
\end{definition}

We recall basic concepts of context-free and regular
languages~\cite{Sipser2013,Mateescu1997}.  A \emph{context-free grammar} is a
tuple $G=(V, \Sigma, R, S)$, where $V$ is the set of nonterminal symbols,
$\Sigma$ is the set of terminals, $R\subseteq V\times(V\cup\Sigma)*$ is the set
of production rules, and $S\in V$ is the start symbol. $G$ is \textit{linear} if
every production rule is of the form $A\rightarrow uBv$ or $A\rightarrow u$,
where $u, v\in\Sigma^*$ and $A, B\in V$. $G$ is right-linear if every production
rule is of the form $A\rightarrow uB$ or $A\rightarrow u$, where
$u\in\Sigma\cup\{\varepsilon\}$ and $A\in V$. The class of linear languages
{\sffamily LIN} consists of languages generated by linear grammars. The class of
regular languages {\sffamily REG} consists of languages generated by
right-linear grammars.

\section{Folding over regular languages}
\label{regreg}

We consider languages of the class $\mathcal{F}(\text{\sffamily REG}, \text{\sffamily REG})$. First, we show that $\mathcal{F}(\text{\sffamily REG}, \text{\sffamily REG})\subseteq \text{\sffamily LIN}$, where {\sffamily LIN} is the class of linear languages.

\begin{theorem}
The class of languages generated by F-systems with regular core and procedure languages is a subset of the class of linear languages.
\label{th1}
\end{theorem}

\begin{proof}
Consider the F-system $\Phi=(L_1, L_2)$ with $L_1, L_2 \in \text{\sffamily REG}$. 
Let $G_1=(V_1, \Sigma, R_1, S_1)$ and $G_2=(V_2, \Gamma, R_2, S_2)$ be right-linear grammars for languages $L_1^\mathcal{R}$ and $L_2^\mathcal{R}$, respectively. Then, a linear grammar $G=(V, \Sigma, R, S)$ for $L(\Phi)$ may be obtained by letting:
\begin{enumerate}
\item $V=V_1\times V_2$,
\item $R=R_\str{u} \cup R_\str{d} \cup R_\eee$, where 
\begin{align*}
R_\str{u}&=\{(A, B)\rightarrow a(C, D)\mid A\rightarrow aC \in R_1, B\rightarrow \str{u}D\in R_2\},\\ 
R_\str{d}&= \{(A, B)\rightarrow (C, D)a\mid A\rightarrow aC \in R_1, B\rightarrow \str{d}D\in R_2\},\\ 
R_\eee &= \{(A, B)\rightarrow \eee \mid A\rightarrow \eee \in R_1, B\rightarrow \eee \in R_2\},
\end{align*}
\item $S=(S_1, S_2)$.
\end{enumerate}

Now, for any nonterminal $A$ of a grammar $G$, let $G^A$ denote the version of $G$ with $A$ as the start symbol. With a straightforward inductive argument we prove the following claim. 
\begin{claim*}
\label{claim1}
For any $A_1\in V_1$ and $A_2 \in V_2$, $L(L({G_1^{A_1}})^\mathcal{R},L({G_2^{A_2}})^\mathcal{R}) = L(G^{(A_1,A_2)})$.
\end{claim*}
\noindent
Naturally, the above claim proves that $L(\Phi)= L(G)$.
\end{proof}

\begin{example}
  \label{ex1}
  Let $\Phi=(L_1,L_2)$ with $L_1= (abc)^*$ and $L_2^*=(\str{udd})^*$, and take the
  following right-linear grammars $G_1$ and $G_2$ defining
  $L_1^\mathcal{R} = (cba)^*$ and $L_2^\mathcal{R}=(\str{ddu})^*$, respectively.
  \begin{align*}
    G_1:{}
    & S_0 \to \epsilon \vvv cS_1&
    & S_1 \to b S_2 &
    & S_2 \to a S_0\\
    G_2 : {}
    & T_0 \to \epsilon \vvv \str{d}T_1&
    & T_1 \to \str{d} T_2 &
    & T_2 \to \str{u} T_0
  \end{align*}
  The construction in the proof above yields the following linear grammar
  (nonproductive rules omitted).
  \begin{align*}
    G : {}
    & (S_0,T_0) \to \epsilon \vvv (S_1,T_1)c &
    & (S_1,T_1) \to (S_2,T_2) b &
    & (S_2,T_2) \to a (S_0,T_0)
  \end{align*}
  Clearly, $L(G)=\{a^n(bc)^n \mid n\geq 0\}=L(\Phi)$.
\qed
\end{example}

Now, we show that $\mathcal{F}(\text{\sffamily REG}, \text{\sffamily REG})\neq \text{\sffamily LIN}$. The proof relies on an \emph{interchange} property of languages generated by folding: if $w_1,w_2\in L(\Phi)$, $|w_1|=|w_2|$, $w_1=h(v_1,u_1)$, and $w_2=h(v_2,u_2)$, then $h(v_1,u_2)$ also belongs to $L(\Phi)$. We construct a linear language that does not have this property.

\begin{theorem}
The class of languages generated by F-systems with regular core and procedure languages is not equal to the class of linear languages.
\label{th2}
\end{theorem}

\begin{proof}
We present a linear language $L$ that cannot be generated by any folding system with regular core and regular procedure languages. The language $L$ over the alphabet $\Sigma=\{a, b, c, d, e, f, \#\}$ is defined with the following linear grammar:
\[
G: S \to S_1 \vvv S_2, \qquad  
S_1 \to a S_1 bc \vvv a\#bc, \qquad 
S_2 \to de S_2 f \vvv de\#f.
\]
Suppose now that there is an F-system $\Phi=(L_1, L_2)$, such that $L=L(\Phi)$, and let $N_1$ and $N_2$ be the numbers of nonterminals of the right-linear grammars that define $L_1$ and $L_2$, respectively. We point out that $L$ has only words of length $3i+1$ for $i\geq1$, and without loss of generality, we assume that both $L_1$ and $L_2$ have words of length $3i+1$ only.
Otherwise, we can take their intersections with the respective regular languages of words of length $3i+1$.

Since every word in $L$ has exactly one occurrence of $\#$, so does every word in $L_1$. Moreover, with a pumping argument we show that $\#$ is in the beginning of every word in $L_1$. More precisely, we let $N=N_1N_2$ and make the following claim. 
\begin{claim*}
\label{claim2}
For every word $w\in L_1$, the symbol $\#$ is present in the first $N$ symbols of $w$.
\end{claim*}

Next, let $n=2N$ and take the words $w \in L_1$ and $v \in L_2$ such that $h(w,v) = a^n{\#}(bc)^n$. Note that $|w| = |v|=6N+1$. Let $w=w_1\#w_2$ and observe that since $|w_1|< N$, $w_2$ contains more than $3N$ symbols in $\{b,c\}$. Because those symbols follow \#, they must be folded down, and therefore $v$ must also contain at least $3N+1$ occurrences of $\str{d}$.

Now, take the words $w' \in L_1$ and $v' \in L_2$ such that $h(w',v') = (de)^n\#f^n$, and consider folding $w'$ according to $v$ ($w'$ and $v$ have the same length). Because $w'$ contains only symbols in $\{d,e,f,\#\}$ the result $h(w',v)$ must also be equal to $(de)^n\#f^n$ ($L$ demands it). However, we observe that $w'=w_1'\#w_2'$ and $|w_1'\#| \leq N$, and therefore, at least $2N+1$ symbols of $w_2'$ is folded down by $v$. Consequently, the result $h(w',v')$ has more than $n$ symbols following $\#$, which contradicts $h(w',v) = (de)^n\#f^n$.
\end{proof}

\section{Conclusion}

From Theorems \ref{th1} and \ref{th2}, we conclude that $\mathcal{F}(\text{\sffamily REG}, \text{\sffamily REG})\subset \text{\sffamily LIN}$. It is also known that $\text{\sffamily REG}\subset\mathcal{F}(\text{\sffamily REG}, \text{\sffamily REG})$ \cite{Sburlan2011}, which places $\mathcal{F}(\text{\sffamily REG}, \text{\sffamily REG})$ as an intermediate class between the regular and linear languages. Interestingly, Theorem~\ref{th2} also shows that $\mathcal{F}(\text{\sffamily REG}, \text{\sffamily REG})$ is not closed under union: the linear language $L$ used in the proof is the union of $L((abc)^*,(\str{udd})^*)$ and $L((edf)^*,(\str{uud})^*)$. Tackling the questions of closure under intersection and complement would require dedicated tools and we leave it for future work. 

A previous work \cite{Lucero2019} introduced a weak pumping lemma stating conditions for a language to belong to $\mathcal{F}(\text{\sffamily REG}, \text{\sffamily REG})$. However, the present result implies that the class must also satisfy the pumping lemma for linear languages \cite{Autebert1997, Horvath2010}, which has stronger conditions than the previous lemma. The relation of the class with the linear languages also implies that it has efficient recognition algorithms of $\mathcal{O}(n^2)$ time and $\mathcal{O}(n)$ space complexities \cite{Kutrib2008}, which may be relevant for applications in natural computing. 

It is also interesting to note that F-systems may be expressed in terms of families of permutations as defined in \cite{Fernau2000}. Since the even-linear languages~\cite{Amar1964}, generated by linear grammars with rules $S\to u S' v$ such that $|u|=|v|$, may be obtained from permutations on regular languages \cite[Example 9]{Fernau2000}, then this class is contained within $\mathcal{F}(\text{\sffamily REG}, \text{\sffamily REG})$.

\section*{Acknowledgments}
We are grateful to the anonymous reviewer who pointed out the relation of $\mathcal{F}(\text{\sffamily REG}, \text{\sffamily REG})$ with the class of even-linear languages.  Jorge C. Lucero was supported by a grant from the Deans of Research and Innovation and of Graduate Studies of the University of Brasília.

\bibliographystyle{elsarticle-num} 
\bibliography{fsystem}

\appendix

\section{Proof of Claim in Theorem \ref{th1}}
\label{a1}

\begin{proof}
First, we show that any word $s\in L(L({G_1^{A_1}})^\mathcal{R},L({G_2^{A_2}})^\mathcal{R})$ is also in $L(G^{(A_1,A_2)})$. 
If $s\in L(L({G_1^{A_1}})^\mathcal{R},L({G_2^{A_2}})^\mathcal{R})$, then there are words $w\in L({G_1^{A_1}})^\mathcal{R}$ and $v\in L({G_2^{A_2}})^\mathcal{R}$ such that $|s|=|w|=|v|$ and $s=h(w, v)$, where $h$ is the folding function defined in Definition~\ref{def1}. Using induction on the length of $s$:

\begin{enumerate}

\item If $|s|=0$, then $s=w=v=\eee$, and $G_1$ and $G_2$ have rules $A_1\to \eee$ and $A_2\to\eee$, respectively. Therefore, $G$ has the rule $(A_1, A_2) \to \eee$, and $\eee\in L(G^{(A_1, A_2)})$.  

\medskip

\item If $|s|>0$, then let $w=w'a$, $v=v'b$, where $a\in\Sigma$ and $b\in\Gamma$. Since $w^\mathcal{R}=a{w'}^\mathcal{R}$ and $v^\mathcal{R}=b{v'}^\mathcal{R}$, then $G_1$ and $G_2$ have rules $A_1\to aB_1$ and $A_2\to bB_2$, respectively, where ${w'}^\mathcal{R}\in L({G_1^{B_1}})$ and ${v'}^\mathcal{R}\in L({G_2^{B_2}})$. Therefore, $G$ has either the rule $(A_1, A_2) \to a(B_1, B_2)$, if $b=\str{u}$, or the rule $(A_1, A_2) \to (B_1, B_2)a$, if $b=\str{d}$.

\medskip

Assume, by induction hypothesis, that $h(w', v')\in L(G^{(B_1, B_2)})$. If $b=\str{u}$, then $(A_1, A_2)$ generates $ah(w',v')=h(w'a, v'\str{u})=h(w,v)$. If $b=\str{d}$, then $(A_1, A_2)$ generates $h(w',v')a=h(w'a, v'\str{d})=h(w,v)$. In either case, $s=h(w,v)\in L(G^{(A_1, A_2)})$.

\end{enumerate}

Next, we show that any word $s\in L(G^{(A_1,A_2)})$ is also in \linebreak $L(L(G_1^{A_1})^\mathcal{R}, L(G_2^{A_2})^\mathcal{R})$. Again, using induction on the length of $s$:

\begin{enumerate}

\item If $|s|=0$, then $s=\eee$ and $G$ has a rule $(A_1, A_2)\rightarrow\eee$. Therefore, $G_1$ and $G_2$ have rules $A_1\to \eee$ and $A_2\to\eee$, respectively, and $h(\eee, \eee) =\eee\in  L(L(G_1^{A_1})^\mathcal{R},L(G_2^{A_2})^\mathcal{R})$. 

\medskip

\item If $|s| > 0$, then $G$ has either a rule $(A_1, A_2) \to a(B_1, B_2)$ or a rule $(A_1, A_2) \to (B_1, B_2)a$, where $a\in\Sigma$. Consider the former case, and
let $s=as'$, where $s'\in L(G^{(B_1, B_2)})$. 

\medskip

By induction hypothesis, assume that $s'\in L(L(G_1^{B_1})^\mathcal{R}, L(G_2^{B_2})^\mathcal{R})$. Then, there are words $w'\in L(G_1^{B_1})^\mathcal{R}$ and  $v'\in L(G_2^{B_2})^\mathcal{R}$ such that $s'=h(w', v')$. Also,  rule $(A_1, A_2) \to a(B_1, B_2)$ implies that $G_1$ and $G_2$ have rules $A_1\to aB_1$ and $A_2\to \str{u}B_2$, respectively, and then $w'a\in L(G_1^{A_1})^\mathcal{R}$ and $v'\str{u}\in L(G_2^{A_2})^\mathcal{R}$. Thus, $h(w'a, v'\str{u})=ah(w',v')=s \in L(L(G_1^{A_1})^\mathcal{R},L(G_2^{A_2})^\mathcal{R})$.

\medskip

The case in which $G$ has a rule $(A_1, A_2) \to (B_1, B_2)a$ is treated similarly, with $s=s'a$. We obtain that $G_2$ has a rule $A_2\to \str{d}B_2$, and then $v'\str{d}\in L(G_2^{A_2})^\mathcal{R}$. Thus, $h(w'a, v'\str{d})=h(w',v')a=s \in L(L(G_1^{A_1})^\mathcal{R},L(G_2^{A_2})^\mathcal{R})$.
\end{enumerate}
\end{proof}

\section{Proof of Claim in Theorem \ref{th2}}
\label{a2}

\begin{proof}
Suppose 
that there is a word $w\in L_1$ such that $w=w_1\# w_2$ such that $|w_1| > N$, take any $v\in L_2$ such that $h(w,v)=u_1\# u_2$. Now let $v=v_1 v_2$ with $|v_1|=|w_1|$. Since $|v_1|=|w_1|> N$, there are $x_1,x_2,y_1,y_2,z_1,z_2$ such that  $w_1=x_1y_1z_1$, $v_1=x_2y_2z_2$, $|x_1|=|x_2|$, $|y_1|=|y_2|>0$, $|z_1|=|z_2|$, and $x_1y_1^kz_1\#w_2\in L_1$ and $x_2y_2^kz_2v_2\in L_2$.  It is straightforward to see that fact if we view the grammars that define $L_1$ and $L_2$ as finite automata, with number of states $N_1$ and $N_2$, respectively. Assign to every position in $w_1$ and $v_1$ the pair of states reached by the two automata when reading those words. Since $|w_1|=|v_1|>N$, then some pair of states appears at least twice, and the positions in $w_1$ and $v_1$ where the same pair has appeared delimit $y_1$ and $y_2$. The rest follows from the pumping lemma for regular languages.

Next, let $m=2|w_2|+1$ and take the corresponding words  $w'=x_1y_1^mz_1\#w_2$ and $v'=x_2y_2^mz_2v_2$. Note that when folding $w'$ under the control of $v'$, before $\#$ is reached the prefix $x_1y_1^mz_1$ is folded into a word $u'$ of length greater than $2|w_2|+1$. Regardless of how the remaining part of $w'$ is folded, the end result will be a word $u_1\# u_2$ where one of $u_1$ or $u_2$ contains the factor $u'$ and therefore is longer than $2|w_2|+1$ while the other word is no longer than $|w_2|$. This however contradicts the observation that follows immediately from the definition of $L$: every word of $L$ has the form $u_1\# u_2$ such that $|u_1| \leq 2 |u_2|$ and $|u_2| \leq 2|u_1|$. 
\end{proof}

\end{document}